\documentclass[colorlinks, aps, prd, amsmath, amssymb, floats, floatfix, nofootinbib, superscriptaddress, showpacs, notitlepage]{revtex4-1}
\usepackage[utf8]{inputenc}
\usepackage{color}
\usepackage{graphicx}
\usepackage{fancyhdr}
\usepackage{enumitem}
\usepackage{hyperref}
\hypersetup{
    colorlinks,	
    citecolor=blue,
}
\newcommand{\be}{\begin{eqnarray}}
\newcommand{\ee}{\end{eqnarray}}

\begin{document}
\title{Iron line spectroscopy with Einstein-dilaton-Gauss-Bonnet black holes}

\author{Sourabh~Nampalliwar}
\email[Corresponding author: ]{sourabh.nampalliwar@uni-tuebingen.de}
\affiliation{Theoretical Astrophysics, Eberhard-Karls Universit\"at T\"ubingen, 72076 T\"ubingen, Germany}

\author{Cosimo~Bambi}
\affiliation{Center for Field Theory and Particle Physics and Department of Physics, Fudan University, 200433 Shanghai, China}
\affiliation{Theoretical Astrophysics, Eberhard-Karls Universit\"at T\"ubingen, 72076 T\"ubingen, Germany}

\author{Kostas~D.~Kokkotas}
\affiliation{Theoretical Astrophysics, Eberhard-Karls Universit\"at T\"ubingen, 72076 T\"ubingen, Germany}

\author{Roman~A.~Konoplya}
\affiliation{Theoretical Astrophysics, Eberhard-Karls Universit\"at T\"ubingen, 72076 T\"ubingen, Germany}
\affiliation{Institute of Physics and Research Centre of Theoretical Physics and Astrophysics, Faculty of Philosophy and Science, Silesian University in Opava, Opava, Czech Republic}
\affiliation{Peoples Friendship University of Russia (RUDN University), 6 Miklukho-Maklaya Street, Moscow 117198, Russian Federation}


\date{\today}

\begin{abstract}
Einstein-dilaton-Gauss-Bonnet gravity is a well-motivated alternative theory of gravity that emerges naturally from string theory. While black hole solutions have been known in this theory in numerical form for a while, an approximate analytical metric was obtained recently by some of us, which allows for faster and more detailed analysis. Here we test the accuracy of the analytical metric in the context of X-ray reflection spectroscopy. We analyze innermost stable circular orbits (ISCO) and relativistically broadened iron lines and find that both the ISCO and iron lines are determined sufficiently accurately up to the limit of the approximation. We also find that, though the ISCO increases by about $7\%$ as dilaton coupling increases from zero to extremal values, the redshift at ISCO changes by less than $1\%$. Consequently, the shape of the iron line is much less sensitive to the dilaton charge than expected.
\end{abstract}

\maketitle
\nopagebreak
\section{Introduction}
Einstein's theory of gravity has been the standard framework for describing gravitational effects in our universe. Since its proposition, it has been applied quite successfully in various astrophysical scenarios. Its predictions have largely been validated in the so called weak field regime~\cite{Will2014}, whereas in the strong field regime it is largely untested. Tests in strong field gravity are becoming more accessible and popular with latest technology. As one of the most compact objects predicted by general relativity, black holes are natural laboratories for testing strong gravity. Within general relativity, most black holes are expected to be described by the uncharged and rotating metric discovered by Kerr~\cite{Kerr1963}.\footnote{There are additional assumptions like four dimensions, aymptotic flatness, etc. See, e.g.,~\cite{Chrusciel2012}.} Besides charge, which is expected to be extremely small for these objects, all the deviations from a Kerr solution are quickly radiated away~\cite{Price1971,Price1972} and the no-hair theorem~\cite{Carter1971,Robinson1975} holds for these objects.

Despite its successes, there are some fundamental questions, e.g., dark matter and dark energy, that are unresolved within Einstein's theory. Moreover, combining Einstein's theory with quantum mechanics results in a non-renormalizable effective theory, which breaks down at the Planck scale. This remains an outstanding problem in physics and a number of alternative theories have been proposed to resolve these issues.
One of the most interesting alternatives is the {\it Einstein-dilaton-Gauss-Bonnet} (EdGB hereafter) theory. It has an additional (to Einstein's theory) term in the action which is second-order in curvature, known as the Gauss-Bonnet term, and is coupled to a dynamical scalar field. This model emerges naturally in string theory where the scalar field is the dilaton~\cite{Kanti1995,Kanti1997}. Black hole solutions in numerical form are known in this theory, for spherically symmetric ~\cite{Kanti1995} as well as rotating cases~\cite{Kleihaus2011,Kleihaus2015}. (See also perturbative solutions at~\cite{Ayzenberg2014,Maselli2015}.)
Various potentially observable properties of the EdGB black hole have been recently studied in a number of works. Quasinormal modes were computed in \cite{Blazquez-Salcedo:2017txk}, while the shadows were found first in \cite{Younsi:2016azx} for the perturbative solution and in \cite{Cunha2016} for the numerical one.

A promising technique for probing the strong field region of black holes is X-ray reflection spectroscopy. The standard approach to analyze black holes with this technique is the disk-corona model~\cite{martocchia1996}. In this model, the black hole is surrounded by a geometrically thin and optically thick disk~\cite{Novikov1973} with accreting matter and possesses a ``corona". The disk is formed of material either from a companion star, in case of stellar-mass black holes in binary systems, or the interstellar medium, in case of supermassive black holes at galactic centers. The disk emits like a blackbody locally, and as a multi-temperature blackbody when integrated radially. The temperature of the inner part of the accretion disk typically is in the soft X-ray band (0.1 -- 1 keV) for stellar-mass black holes and in the optical/UV band (1 -- 100 eV) for the supermassive ones. The corona is a hotter ($\sim$100 keV) and optically thin source near the black hole. The morphology of the corona is not very well understood. (See, e.g.,~\cite{wilkins2012,wilkins2014}.) Thermal photons from the disk gain energy via inverse Compton scattering off the hot electrons in the corona, and transform into X-rays with a characteristic power-law distribution. These reprocessed photons return to the disk, producing a reflection component with fluorescent emission lines. The strongest feature of the reflection spectrum is the iron K$\alpha$ line, since the disk is usually abundant in iron and the fluorescent yield for iron is higher than lighter elements, with emission lines at 6.4 keV in the case of neutral or weakly ionized iron but can go up to 6.97 keV for H-like ions.

While the iron K$\alpha$ line is a narrow line in the rest-frame of the disk, relativistic effects due to the gravity of the central black hole cause this {\it line} to broaden and skew for observers far away. Combination of all such broadened lines, from different ionizations of iron as well as from other elements present in the accretion disk, produces the {\it reflection spectrum}. With high quality observations and suitable model of the disk, corona, etc., analysis of the reflection spectrum can be a powerful tool for probing the strong gravitational fields of accreting black holes~\cite{Bambi2015,relxillnk,Cao2017}. For the rest of this paper, we focus our attention on the iron line, since the phenomenologies of a single line and the complete reflection spectrum are similar.

Iron line spectroscopy was first applied to EdGB black hole metrics in~\cite{Zhang2017}. They used the numerical metric of~\cite{Kleihaus2011,Kleihaus2015} to
simulate observations of iron lines with current (NuSTAR\footnote{\url{https://www.nustar.caltech.edu/}}) and future (LAD/eXTP~\cite{Zhang2016}) instruments. They tried to recover the input parameters with the standard Kerr iron line data analysis model RELLINE~\cite{Dauser2014}. The logic behind this approach is as follows: a good fit with RELLINE precludes the possibility of detecting non-Kerr metrics (the EdGB black hole metric in this case) with iron line spectroscopy. If a good fit is not possible, it suggests that the non-Kerr metric sufficiently alters the iron line to make this technique a useful approach for testing this non-Kerr metric. They found some unresolved features in LAD/eXTP simulations which could not be fitted with a Kerr model. This suggests that X-ray spectroscopy in near future would be able to test EdGB black hole metrics with real observations.

For this proof-of-principle study a numerical metric sufficed, but there are various drawbacks in using such metrics:
\begin{enumerate}
\item Calculating propagation of photons along geodesics is relatively slower in numerical metrics, since metric coefficients and Christoffel symbols need to be calculated through interpolation.
\item Errors due to interpolation require delicate handling to ensure they are within acceptable limits.
\item Pathologies may appear in non-Kerr metrics which would not be apparent if the metric is available only in a numerical form.
\end{enumerate}
While a numerical metric was sufficient to claim that the EdGB black holes would have observational signatures distinct from Kerr, to quantify the differences and develop a model that can calculate the differences with real observational data, it is crucial to have analytical expressions for the metric. Recently, some of us obtained an approximate analytical metric for the spherically symmetric EdGB black holes~\cite{Kokkotas2017}, based on the continued fraction expansion of~\cite{Konoplya2016}. The expressions are relatively compact and provide excellent accuracy for the metric components.

In the present work we test the accuracy of the approximate analytical metric for X-ray reflection spectroscopy. We compare the radius of the innermost stable circular orbit (ISCO), which usually determines the inner edge of the accretion disk and has a strong effect on the low energy part of the iron line, calculated with the numerical metric and the analytical metric. We then compare the iron line with numerical and analytical metrics and show that the analytical metric can produce the iron line with sufficient accuracy. We also discuss an interesting feature where although the dilaton charge changes the radius of the ISCO, the change in the shape of the iron line is much weaker than expected.

The paper is organized as follows: In Section~\ref{sec:edgbbh}, we review the numerical and analytical black hole metrics in EdGB theory. In Section~\ref{sec:ironline}, we review the calculation of iron line and compare iron lines calculated with  the numerical and analytical metrics. An interesting feature regarding the effect of the dilaton charge on the iron line is described and explained in Section~\ref{sec:discussion}. Conclusion follows in Section~\ref{sec:conclusion}.
Throughout, we employ units where $c=G=\hbar=1$ and the metric has the signature $(-+++)$.

\section{black hole metric in EDGB theory\label{sec:edgbbh}}
Following~\cite{Kokkotas2017}, the Lagrangian in Einstein-dilaton-Gauss-Bonnet gravity is given as
\be\label{eq:action}
	\mathcal{L} = \frac{1}{2}R - \frac{1}{4} \partial_{\mu}\partial^{\mu} \phi + \frac{\alpha'}{8g^2} e^{\phi} \left (R_{\mu \nu \rho \sigma}R^{\mu \nu \rho \sigma} - 4R_{\mu \nu}R^{\mu \nu} + R^2 \right ),
\ee
where $\phi$ is the dilaton field and $\alpha'$ and $g$ are coupling constants. $\alpha'$ has units of (length)$^2$ while $g$ and $\phi$ are dimensionless. By a conformal rescaling of the dilaton field, $\alpha'/g^2 \rightarrow 1$. To describe non-rotating black holes, a spherically symmetric spacetime is chosen:
\be
	ds^2 = -e^{\Gamma(r)}dt^2 + e^{\Lambda(r)}dr^2 + r^2 \left (d\theta^2 + \sin^2{\theta}\,d\phi^2 \right ).
\label{eq:metric}
\ee
The dilaton field $\phi(r)$ and the metric functions are defined as follows:
\begin{align}
	\phi''(r) &= -\frac{d_1(r,\Lambda,\Gamma,\phi,\phi')\,}{\,d(r,\Lambda,\Gamma,\phi,\phi')},\\
	\Gamma''(r) &=  -\frac{d_2(r,\Lambda,\Gamma,\phi,\phi')\,}{\,d(r,\Lambda,\Gamma,\phi,\phi')},\\
	e^{\Lambda(r)} &= \frac{1}{2} \left ( \sqrt{Q^2 - 6\phi'e^{\phi} \Gamma'} - Q \right ),\\
	\intertext{where}
	Q &\equiv \frac{\phi'^2r^2}{4} - 1 - \left (r + \frac{\phi'^2e^{}\phi}{2}\Gamma' \right ),
\end{align}
while the expressions for $d, d_1$ and $d_2$ can be referred from~\cite{Kanti1995}.
These equations can be solved with the following initial conditions at the event horizon $r_0$:
\begin{align}
	\phi(r_0) &= \phi_0,\\
	\phi'(r_0) &= r_0e^{-\phi_0}\left (\sqrt{1-6\frac{e^{2e^\phi_0}}{r_0^4}}-1 \right), \label{eq:phidrhorz}\\
	\Psi(r_0) &= 1,
\end{align}
where $\Psi(r) \equiv \Gamma'(r)(r-r_0)$ is introduced since $\Gamma'(r)$ diverges as $1/(r-r_0)$ at the event horizon. For the rest of this paper, we fix $r_0=1$.
Black hole solutions are parameterized by a dimensionless parameter $p$, given as
\be
	p = 6e^{2\phi_0}, \qquad \qquad 0\leq p < 1,
\ee
so that $p=0$ corresponds to the Schwarzschild black hole and $p=1$ is the largest possible value of $p$ that ensures Eq.~\eqref{eq:phidrhorz} remains real.

The analytical metric is given in terms of a compactified radial coordinate
\be
	x=1-\frac{r_0}{r}, \qquad \qquad	 0 < x \leq 1,
\label{eq:xcord}
\ee
as
\begin{align}
	e^{\Gamma} \equiv xA(x), \qquad \qquad e^{\frac{\Gamma+\Lambda}{2}} \equiv B(x),
\label{eq:axbx}
\end{align}
where
\begin{align}
	A(x) &= 1 - \epsilon(1-x) + (a_0-\epsilon)(1-x)^2 + \tilde{A}(x)(1-x)^3, \\
	B(x) &= 1 + b_0(1-x) + \tilde{B}(x)(1-x)^2.
\end{align}
 The coefficients $a_0, b_0$ and $\epsilon$ are introduced to match the post-Newtonian asymptotic at infinity. It turns out that
\be
	a_0 = b_0 = 0,\\
	\epsilon \approx \frac{p}{11} - \frac{p^2}{131}. \label{eq:epsilon}
\ee
The functions $\tilde{A}(x)$ and $\tilde{B}(x)$ are given in terms of continued fractions as
\be
	\tilde{A}(x) = \cfrac{a_1}{1+\cfrac{a_2x}{1+\cfrac{a_3x}{1+\cfrac{a_4x}{1+\cdots}}}}, \\
	\tilde{B}(x) = \cfrac{b_1}{1+\cfrac{b_2x}{1+\cfrac{b_3x}{1+\cfrac{b_4x}{1+\cdots}}}}, \\
\ee
where the coefficients $a_1, a_2,\,\dots$ and $b_1, b_2, \,\dots$ are calculated numerically. At third order of expansion, i.e.,
\be
a_4=a_5=\dots=0, \qquad \qquad b_3=b_4=\dots=0,
\ee
the following rational functions of $p$ give a good fit for the remaining coefficients:
\begin{align}
	a_1 &= \frac{5p}{(1-p)(5-3p)}\left (\frac{p^2}{40} + \frac{p}{19}-\frac{1}{13} \right ), \\
	a_2 &= \frac{3-11p}{(1-p)(2-p)}\left (\frac{15p}{19} - \frac{11}{13} \right ), \\
	a_3 &= \frac{1}{1-p} \left (\frac{22}{9} - \frac{5p}{7} \right ), \\
	b_1 &= -\frac{13p}{(1-p)(13-9p)}\left (\frac{p^2}{8} + \frac{5p}{13}+\frac{7}{27} \right ), \\
	b_2 &= -\frac{1}{(1-p)(5-4p)}\left (\frac{19p^2}{12} + \frac{248p}{19}-\frac{151}{10} \right ).
\end{align}
The fit remains accurate up to $p\approx0.97$, beyond which the continued fraction converges slowly. Thus, for a given dilaton charge $p$ and radial coordinate $r$, we can obtain $A(x)$ and $B(x)$ and calculate the metric coefficients using Eq.~\eqref{eq:xcord} and~\eqref{eq:axbx}.

\section{Iron line calculation\label{sec:ironline}}
We model the accretion disk as a Novikov-Thorne~\cite{Novikov1973} type geometrically thin optically thick disk in the equatorial plane of the black hole. The inner edge is assumed to be located at the ISCO, a typical assumption since no stable circular orbits exist for any smaller radii, and the outer disk at some large value, large enough that the shape of the iron line is largely insensitive to the exact value. The particles on the disk follow nearly circular geodesics. The flux received by a distant observer from such a disk around a black hole is given by
\begin{align}
	N (E_{\textrm{obs}}) = \frac{1}{E_{\textrm{obs}}} \int I_{\textrm{obs}}(E_{\textrm{obs}})\, d\Omega_{\textrm{obs}} = \frac{1}{E_{\textrm{obs}}}  \int g^3 I_{e}(E_{e})\, d\Omega_{\textrm{obs}},
\label{eq:flux}
\end{align}
where $E_{\textrm{obs}}$ and $I_{\textrm{obs}}$ are the photon energy and the specific intensity of the radiation respectively as measured by the distant observer, while $E_{e}$ and $I_{e}$ are the photon energy and the specific intensity respectively at the point of emission in the local emitter's frame. $d\Omega_{\textrm{obs}}$ is the area element on the distant observer's plane. $I_{\textrm{obs}} = g^3I_{e}$ follows from the Liouville's theorem, where $g  = E_{\textrm{obs}}/E_{e}$ is the all important redshift factor. The intensity profile $I_e$ depends on the details, like the morphology and the location of the corona, which is not very well understood. The standard assumption is to consider a power-law like emissivity, and we assume
\begin{align}
	I_e \propto \frac{1}{r^3}.
\label{eq:emis}
\end{align}

To calculate the integral in Eq.~\eqref{eq:flux}, we discretize the integral over the observer plane. To this end, we divide the plane of the observer in concentric ellipses, given as
\begin{align}
X_0 (r, \phi) = r_{\textrm{obs}} \cos{\phi}, \qquad\qquad Y_0 (r, \phi) = r_{\textrm{obs}} \sin{\phi} \cos i,
\label{eq:observer}
\end{align}
where ($X_0, Y_0$) are the Cartesian coordinates and ($r_{\textrm{obs}}, \phi$) are the spherical coordinates on the observer place and $i$ is the inclination of the observer relative to the accretion disk. The range of $r_{\textrm{obs}}$
is chosen to ensure enough photons land on every part of the disk, and is discretized in $N_r$ values. $\phi$ varies from $0$ to $2\pi$ in $N_{\phi}$ steps. Photons are traced back in time, from the observer to the point of emission.  Only those photons that land between the inner and the outer edge are included in the integral of Eq.~\eqref{eq:flux}. $N_r$ and $N_{\phi}$ are chosen to be large enough so that the integral changes little for larger values. For each photon, the constants of motion are calculated at its location on the observer plane and the geodesic equations are solved using the ray-tracing method described in~\cite{Bambi2012}, which uses an adaptive step-size fourth-order Runge-Kutta-Nystr\"{o}m algorithm~\cite{lund2009}. Once the point of emission is known, the redshift $g$ can be calculated. $g$ is given by
\begin{align}
	g = \frac{E_{\textrm{obs}}}{E_e} = \frac{\nu_{\rm o}}{\nu_{\rm e}} = \frac{-u^\mu_{\rm o} k_\mu}{-u^\nu_{\rm e} k_\nu},
\label{eq:redshift}
\end{align}
where $u^\mu_{\rm o} = (1,0,0,0)$ is the 4-velocity of the distant observer, $k^\mu$ is the photon four-momentum, and $u^\nu_{\rm e} = u^t_{\rm e} (1,0,0,\Omega)$ is the four-velocity of the particles on the disk. Moreover, for particles on circular geodesics on the equatorial disk,
\begin{align}
	u^t_{\rm e} = \dot{t} = \frac{1}{\sqrt{- g_{tt} - \Omega^2 g_{\phi\phi}}},
\end{align}
where $g_{tt}$ and $g_{\phi\phi}$ are the coefficients of $dt^2$ and $d\phi^2$ respectively in the metric, and $\Omega$ is the angular velocity, given by
\begin{align}
	\Omega = \pm \sqrt{- \left( \frac{\partial_r g_{tt}}{\partial_r g_{\phi\phi}}\right)}.
\label{eq:omega}
\end{align}
Thus
\begin{align}
g = \frac{\sqrt{-g_{tt} - g_{\phi\phi}\Omega^2}}{1 - \lambda \Omega},
\label{eq-utgg}
\end{align}
where $\lambda = -k_{\phi}/k_t$ is a constant of motion along the photon geodesic and is calculated from the initial conditions.

\begin{figure}[h]
\begin{center}
\includegraphics[type=pdf,ext=.pdf,read=.pdf,width=0.49\textwidth]{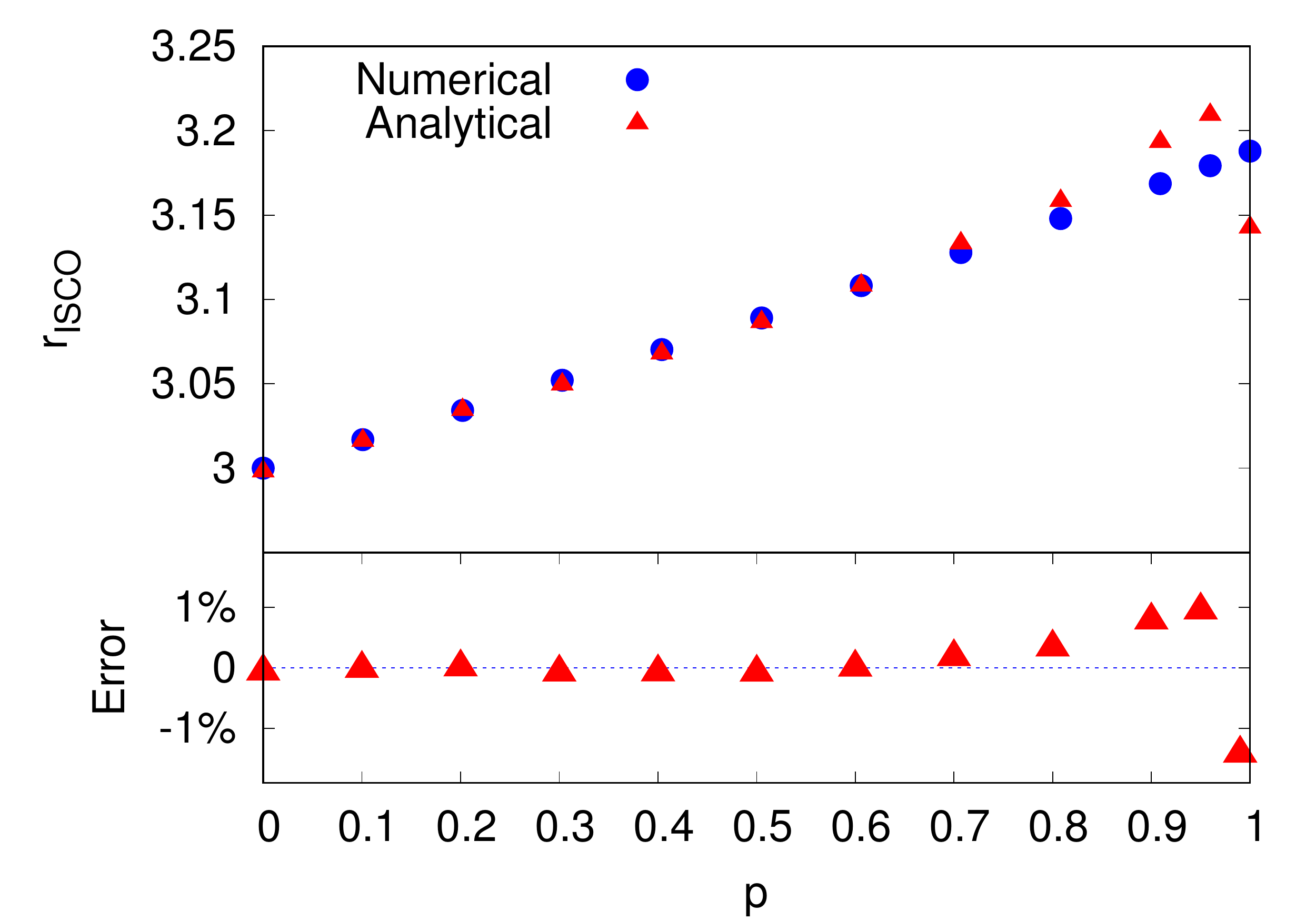}
\end{center}
\caption{{\it Top}: ISCO radii, in units of the horizon radius, plotted for a range of values of $p$ from 0 to 0.99 using the analytical and the numerical EdGB BH metric. {\it Bottom}: Percent error in ISCO radius for same values of $p$ as the top figure. See text for details.\label{fig:isco_comp}}
\end{figure}
One of the crucial factors determining the shape of the iron line is the location of the inner edge, since in the case of Kerr black holes, the region closest to the black hole suffers maximum redshift and generates the low-energy part of the iron line. Therefore, the shape of the iron line is quite sensitive to the location of the inner edge. Since we assume the inner edge to be located at the ISCO, as a first test of the accuracy of the analytical metric we compare the ISCO radius as obtained from the numerical metric and the analytical metric. Details of the calculation of the ISCO radius are given in Appendix~\ref{app:isco}. Fig.~\ref{fig:isco_comp} shows the ISCO radius for several values of $p$ calculated using each metric. Several features are apparent: 
\begin{enumerate}[listparindent=0.7cm]
\item The ISCO radius increases with $p$. 

This can be counter-intuitive. The presence of a scalar field outside the horizon suggests the possibility that some of the energy could be outside the ISCO region. Thus, the ISCO radius in the presence of a non-zero dilaton charge should be smaller, for the same asymptotic mass, than its vacuum counterpart. Eq.~\eqref{eq:riscom_delp} shows that indeed the ISCO radius, in units of the ADM (Arnowitt-Deser-Misner) mass $M$, decreases with dilaton charge. (The expression in that equation is for small $p$, for brevity. For any $p$ the qualitative behavior remains the same. See, e.g.,~\cite{Ayzenberg2014}.) Similarly, and to a much larger extent, the energy stored in the scalar field lies outside the horizon. Therefore, the horizon radius also reduces for a non-zero dilaton charge, for the same asymptotic mass, relative to its vacuum counterpart. Eq.~\eqref{eq:rhorizon} shows the same. While both ISCO radius and horizon radius decrease, since the scalar field is concentrated close to the horizon the effect is much stronger in the case of the horizon radius. Consequently, when calculated in units of the horizon radius, we find that the ISCO radius increases. For small $p$, this relation is given in Eq.~\eqref{eq:riscorh_delp}.

\item The analytical metric can calculate the ISCO radius within 1\% accuracy up to $p=0.97$, which is the limit of the approximation for the analytical metric. 
\item For larger values of $p$, the continued fraction expansion converges slowly and the agreement between the ISCO radii calculated from the numerical and the analytical metrics gets worse.
\end{enumerate}

\begin{figure}[h]
\begin{center}
\includegraphics[type=pdf,ext=.pdf,read=.pdf,width=0.49\textwidth]{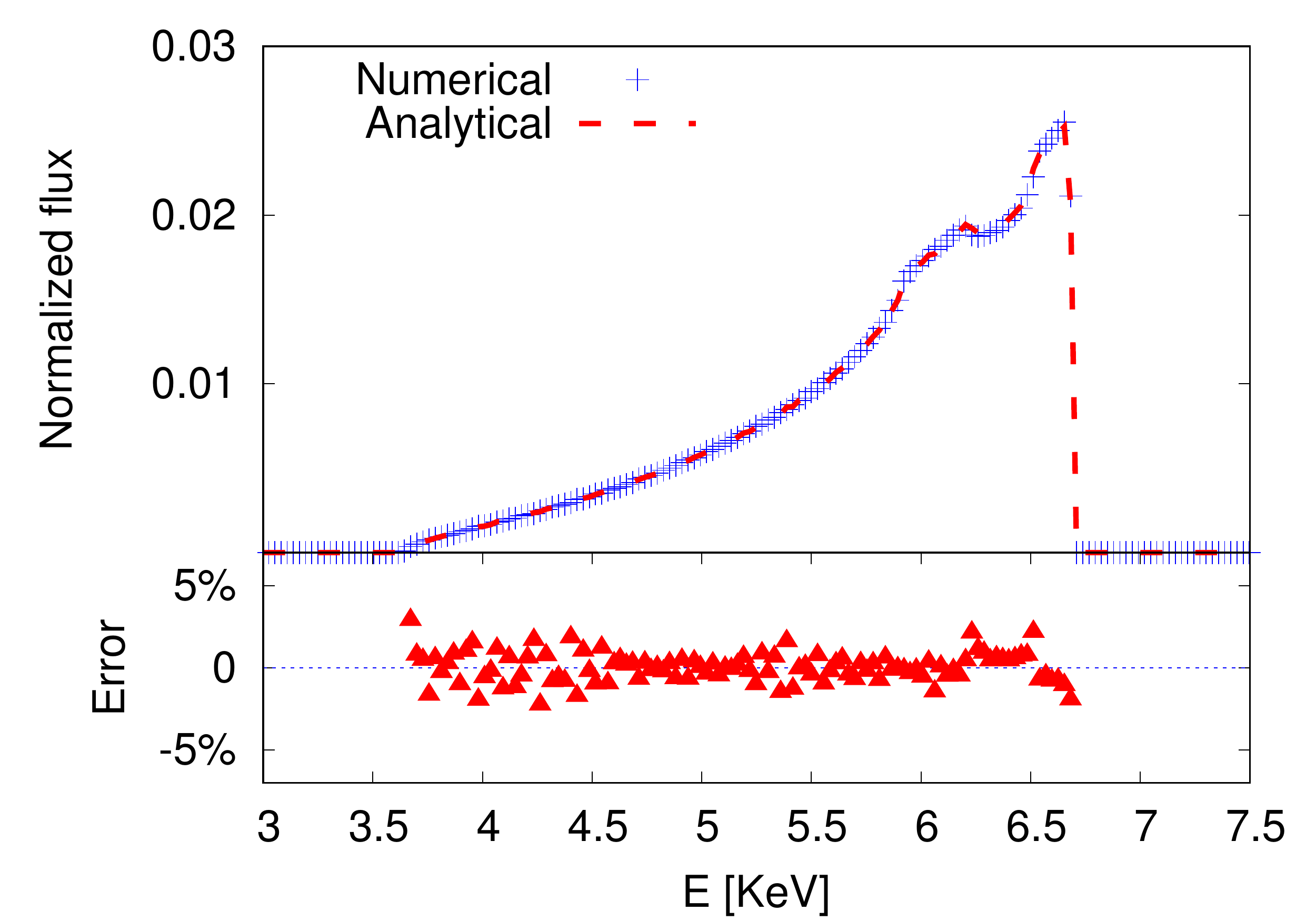}
\includegraphics[type=pdf,ext=.pdf,read=.pdf,width=0.49\textwidth]{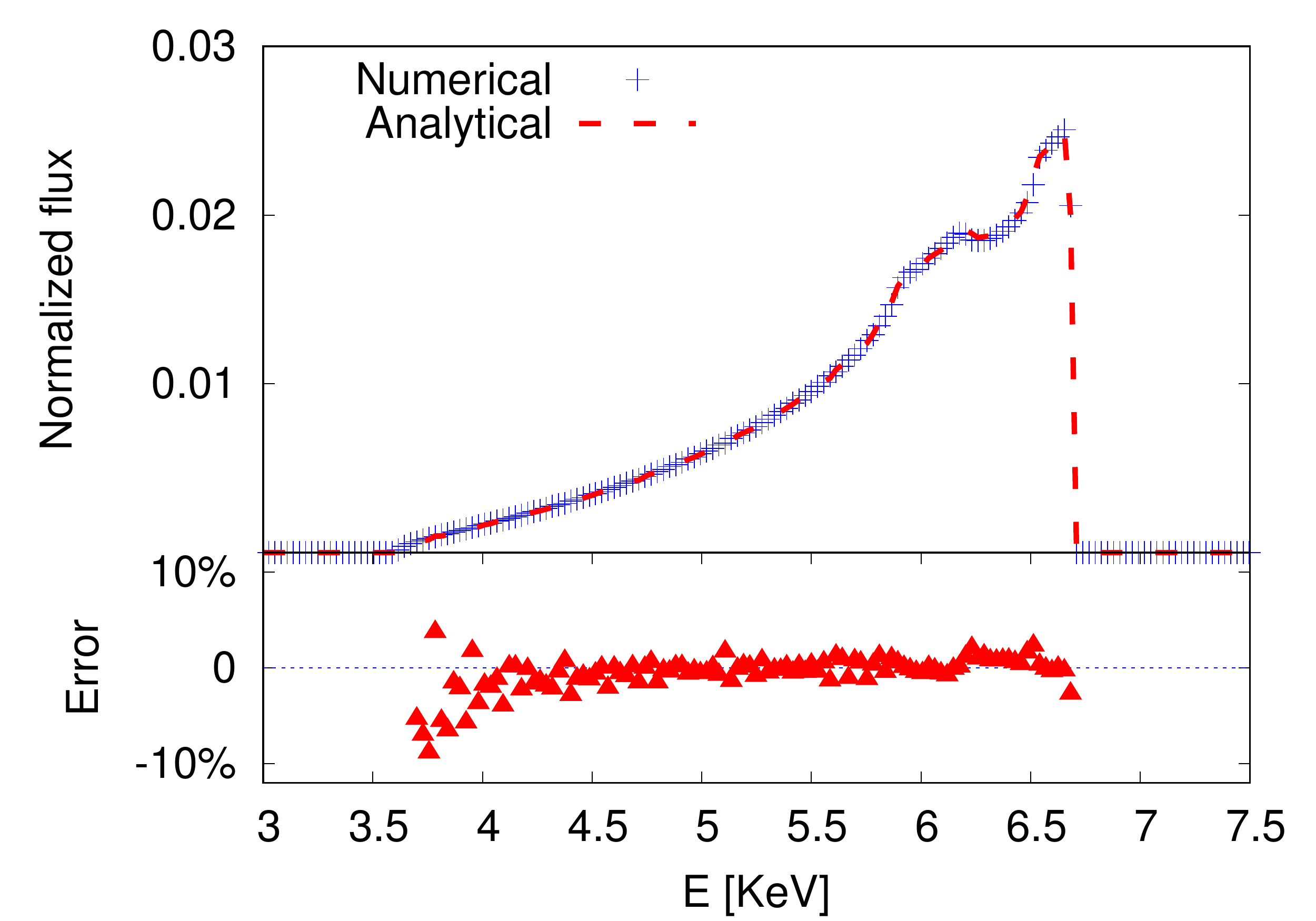}
\end{center}
\caption{Relativistically broadened iron lines for a viewing angle of 30 degrees, calculated using Eq.~\eqref{eq:flux} for: {\it Left}: $p=0$ numerical and $p=0$ analytical EdGB BHs. {\it Right}: $p=0.97$ numerical and $p=0.97$ analytical EdGB BHs. The bottom part in each plot shows the relative percent error. See text for details.\label{fig:lines}}
\end{figure}


We now look at the iron line calculated using the two metrics. The first comparison, shown in Fig.~\ref{fig:lines} in the left plot, is for $p=0$. We see that there is good agreement of both the numerical and the analytical metric iron lines. The percent difference between the two lines is plotted in the lower plot. We can see that the error is of the order of a few percent, and is distributed randomly. This error is essentially acquired during the interpolation of the numerical metric and ray-tracing. After this check of consistency, we now compare the lines for a non-zero $p$. The right plot in Fig.~\ref{fig:lines} shows iron lines calculated with the numerical and the analytical metric respectively, for $p=0.97$. Note that $p=0.97$ is the limit up to which the analytical approximation is reliable, so we expect largest errors in the analytical iron line for this value of $p$. What the figure shows is again good agreement between the lines calculated using the two metrics. The error is again random and can be attributed to computational precision.

\section{Discussion\label{sec:discussion}}
\begin{figure}[h]
\begin{center}
\includegraphics[type=pdf,ext=.pdf,read=.pdf,width=0.49\textwidth]{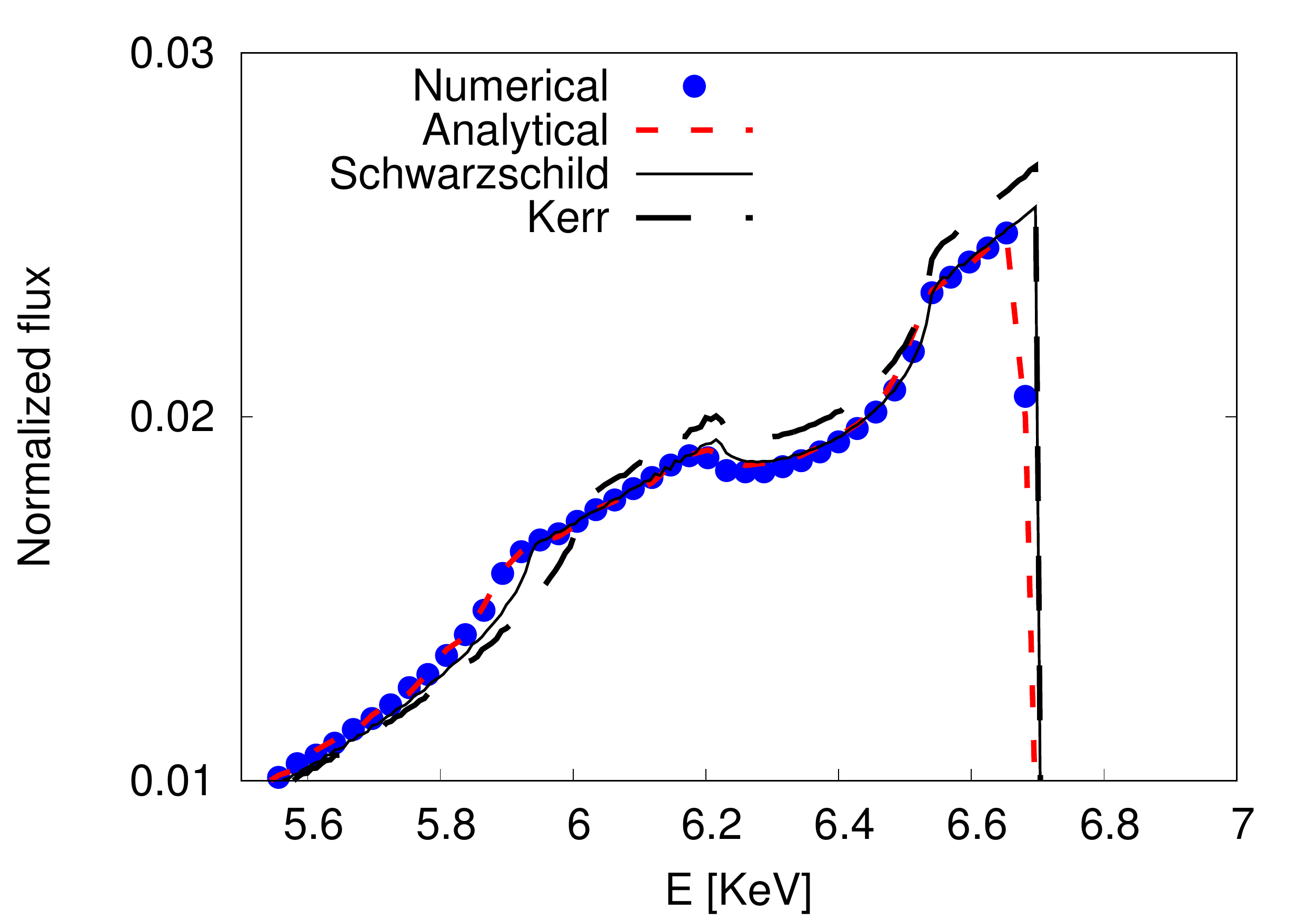}
\includegraphics[type=pdf,ext=.pdf,read=.pdf,width=0.49\textwidth]{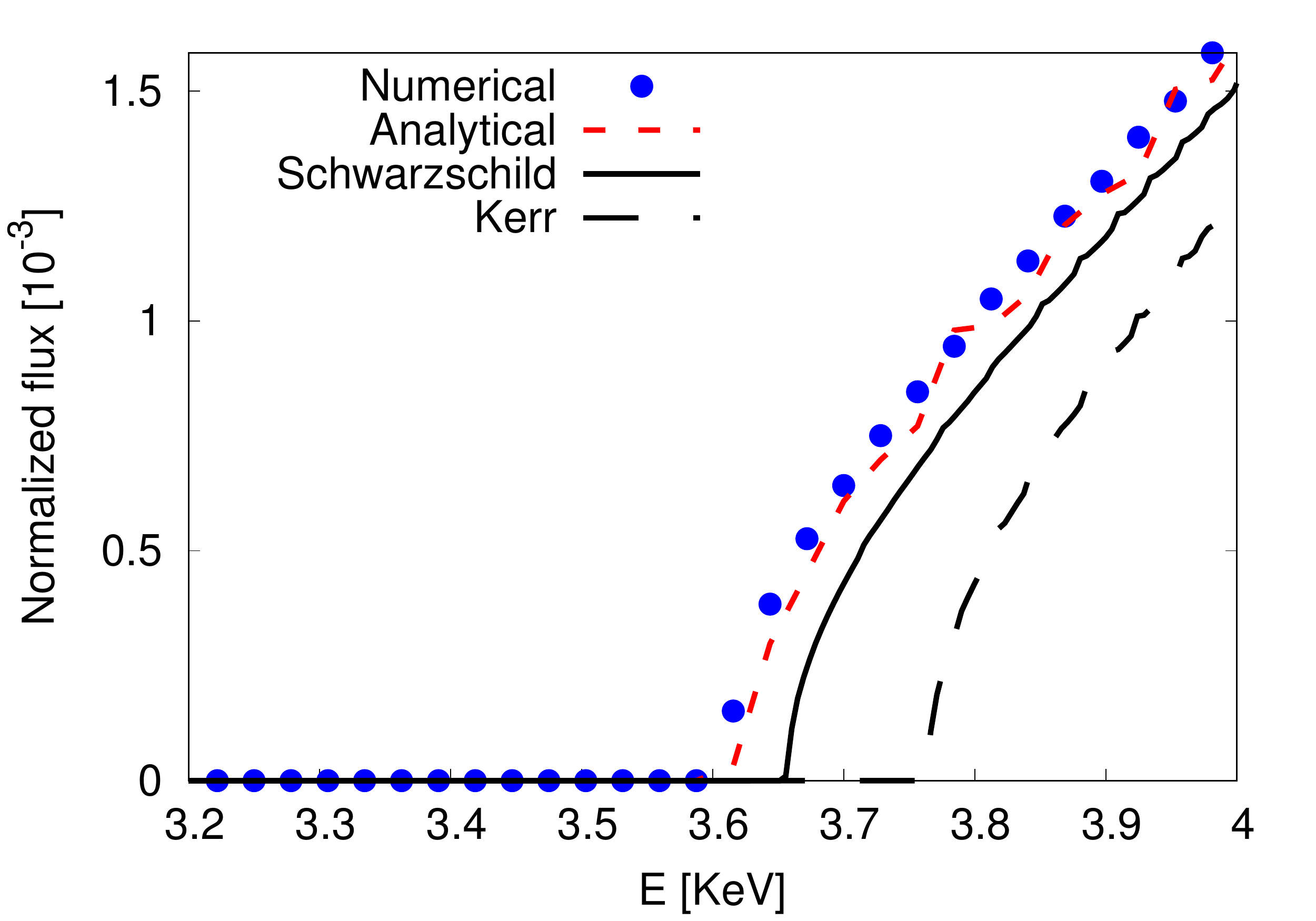}
\end{center}
\caption{Relativistically broadened iron lines for a viewing angle of 30 degrees, calculated using Eq.~\eqref{eq:flux} for $p=0.97$ numerical and $p=0.97$ analytical EdGB BHs, and using RELLINE for a Schwarzschild BH and an $a^*=-0.12$ Kerr BH: {\it Left}: zoomed in the high energy part between $5.5-7$ KeV, {\it Right}: zoomed in the low energy part between $3-4$ KeV. See text for details.\label{fig:lines2}}
\end{figure}
In the previous section, we explored the accuracy of the analytical EdGB black hole metric with iron line spectroscopy. Now we discuss an interesting feature of the iron lines of EdGB black hole metrics. To illustrate the feature, we plot in Fig.~\ref{fig:lines2} the iron lines of $p=0.97$ numerical and $p=0.97$ analytical EdGB BHs with two additional lines: a line with the Schwarzschild metric, and a line with a Kerr metric of spin $a^*=-0.12$.\footnote{A negative $a^*$ means the accretion disk is counter-rotating relative to the black hole.} The ISCO radius of this rotating hole matches with the ISCO radius of the $p=0.97$ EdGB black hole. The viewing angle is the same in all cases. Naively, one would expect the EdGB BH iron lines to be closer to the rotating Kerr iron line, but instead we find that {\it the EdGB BH iron lines are much closer to the Schwarzschild iron line}. 
Since both the numerical and the analytical metric exhibit this feature, besides some strange coincidence, this feature cannot be attributed to either a poor approximation of the analytical metric or a poor interpolation of the numerical metric. Rather, it seems to be a property of the black holes of EdGB theory.

To investigate the peculiar feature of Fig.~\ref{fig:lines2}, we look at the expression for the flux number density in Eq.~\eqref{eq:flux}. For an intensity profile of Eq.~\eqref{eq:emis}, the equation becomes
\begin{align}
	N (E_{\textrm{obs}}) \propto \frac{1}{E_{\textrm{obs}}}  \int \frac{g^3}{r^3} \, d\Omega_{\textrm{obs}}.
\end{align}
We want to focus on the effect of the black hole metric on this integral. The background metric affects the redshift factor and the radial extent of the disk. If we restrict our disk to be very close to the inner edge, we can approximate the above equation as
 \begin{align}
	N_{\textrm{ISCO}}\propto \frac{1}{E_{\textrm{obs}}} \frac{1}{(r_{\textrm{ISCO}})^3}\int (g_{\textrm{ISCO}})^3\, d\Omega_{\textrm{obs}}(r_{\textrm{ISCO}}),
\label{eq:fluxapprox}
\end{align}
where $d\Omega_{\textrm{obs}}(r_{\textrm{ISCO}})$ are those area elements from which a photon lands on the ISCO. Since the low energy part of the iron line is dominated by the radiation from the inner edge, we can use this relation to approximate the behavior of the flux at low energies. Moreover, since the ISCO is a property of the metric, by comparing quantities at ISCO and not at some fixed radial coordinate, we ensure that our inferences are not affected by the choice of the coordinate system.

\begin{figure}[h]
\begin{center}
\includegraphics[type=pdf,ext=.pdf,read=.pdf,width=0.49\textwidth]{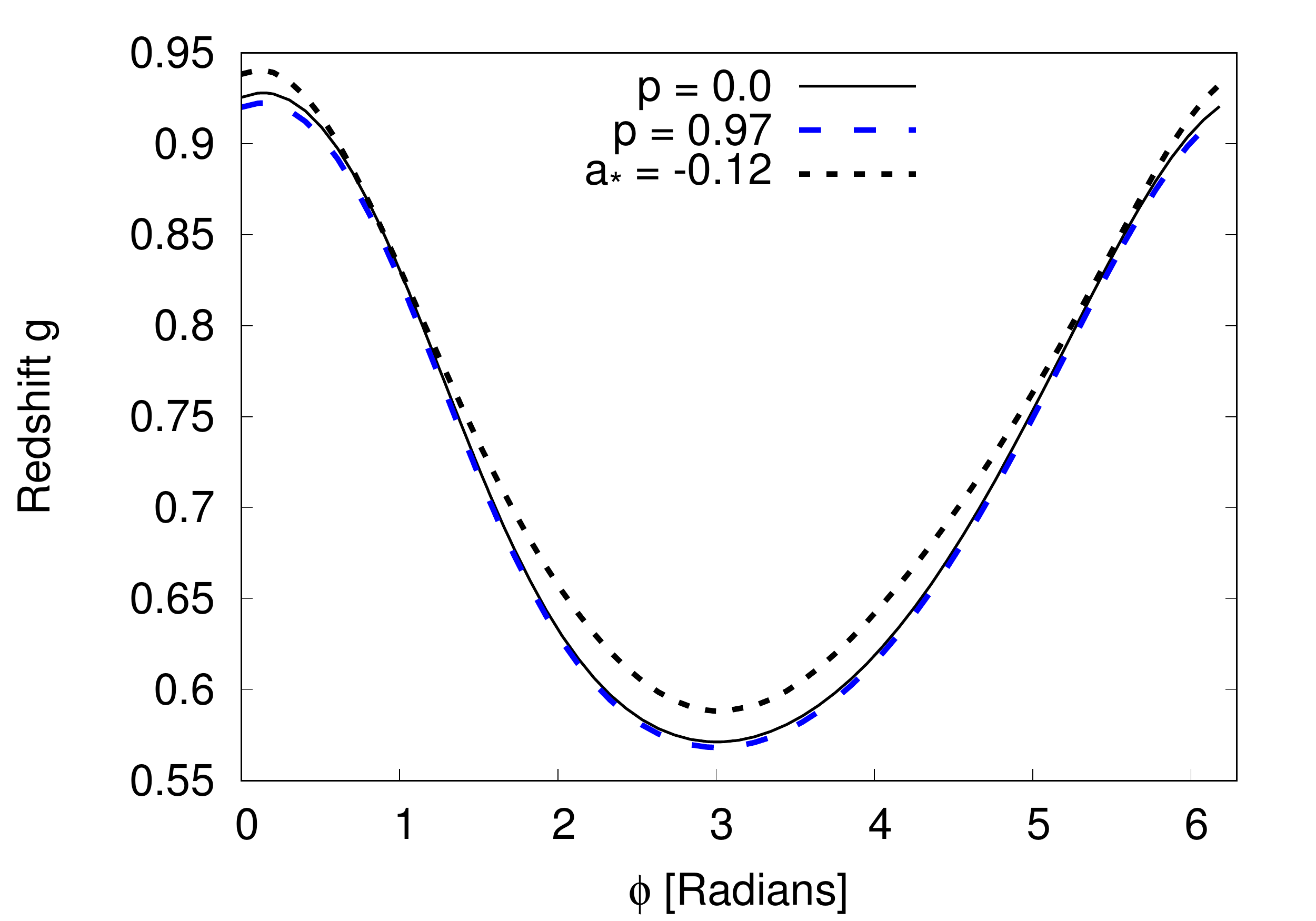}
\includegraphics[type=pdf,ext=.pdf,read=.pdf,width=0.49\textwidth]{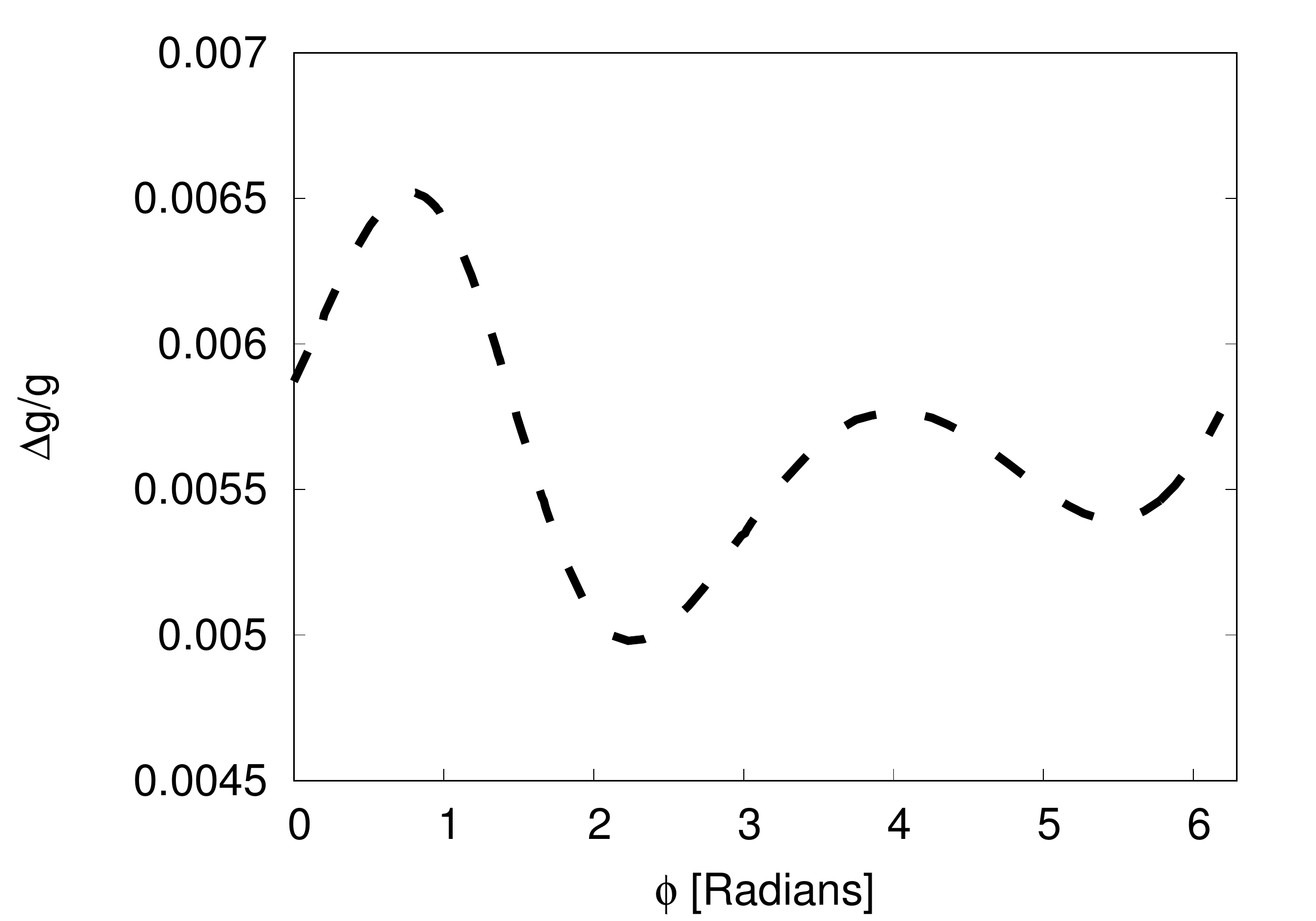}
\end{center}
\caption{{\it Left}: Redshift factor $g = E_{\textrm{obs}}/E_{e}$ calculated for photons emanating from respective ISCOs in three cases: $p=0$ analytical and $p=0.97$ analytical EdGB BHs and $a_*=-0.12$ Kerr BH. {\it Right}: Fractional difference in redshift factor between $p=0$ analytical and $p=0.97$ analytical EdGB BHs, as plotted on the left. See text for details.\label{fig:redshifts}}
\end{figure}
\begin{figure}[h]
\begin{center}
\includegraphics[type=pdf,ext=.pdf,read=.pdf,width=0.49\textwidth]{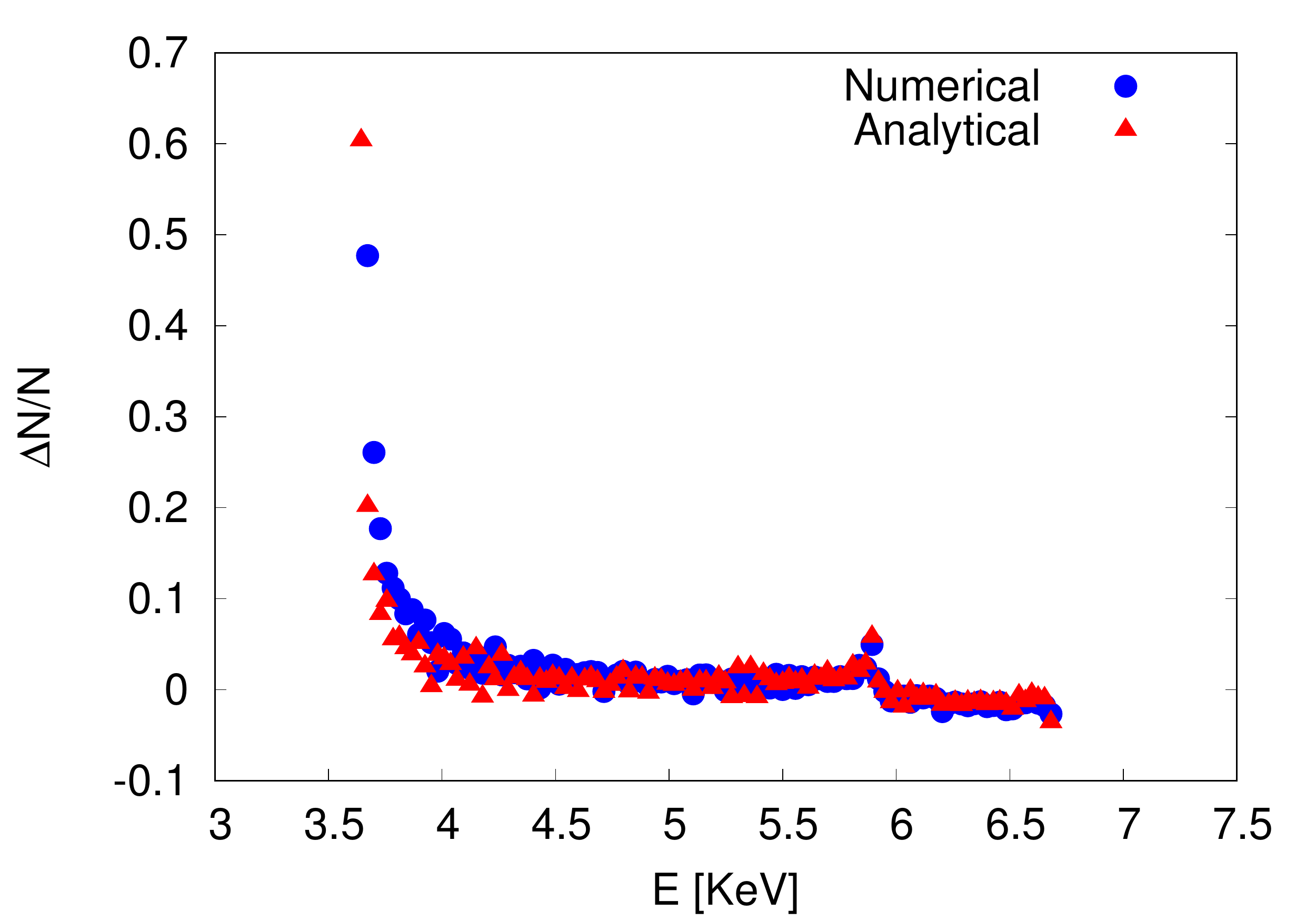}
\end{center}
\caption{Fractional difference between iron lines calculated for $p=0$ analytical and $p=0.97$ analytical EdGB BHs (red crosses) and $p=0$ numerical and $p=0.97$ numerical EdGB BHs (blue plusses), as function of $E_{\textrm{obs}}$. See text for details.\label{fig:lines_diff}}
\end{figure}

Now the question is, how do the redshift factors and the ISCO radii change as we go from $p=0$ to $p=0.97$. We have already seen how $r_{\textrm{ISCO}}$ changes, in Fig~\ref{fig:isco_comp}: there is a $\sim7\%$ increase in $r_{\textrm{ISCO}}$ from $p=0$ to $p=0.97$, for both the numerical and the analytical metric. To study the behavior of $g$, we trace the photon in the same way as described in Sec.~\ref{sec:ironline}, but additionally,  $r_{\textrm{obs}}$ is adjusted in an adaptive way so that photons land precisely at $r_{\textrm{ISCO}}$, after which the value of $g$ can be obtained with Eq.~\eqref{eq:redshift}.\footnote{A similar adaptive approach is used in the calculation of FITS files for non-Kerr metrics, and is described in~\cite{relxillnk}.} The redshift factors of photons associated with $r_{\textrm{ISCO}}$ are shown in Fig.~\ref{fig:redshifts} as a function of their angular position on the observer plane, for $p=0.0$ and $p=0.97$ analytical metric. Also plotted is the redshift for the $a=-0.12$ Kerr line, which has the same ISCO location as a $p=0.97$ EdGB hole. We see that $g$ changes by fractions of a percent as we go from $p=0$ to $p=0.97$. The redshifts in the Kerr case differ much more than the two EdGB cases, by about 15\%.
Since $g_{\textrm{ISCO}}$'s differ by fractions of a percent while $r_{\textrm{ISCO}}$ changes by about ten percent, Eq.~\eqref{eq:fluxapprox} suggests the difference between the $p=0$ and the $p=0.97$ line to be of the order of ten percent at low energies. Fig.~\ref{fig:lines_diff} shows the difference between the two lines and we find that this is indeed the case. The differences are computed in both the analytical and the numerical cases, to ensure that no spurious errors due to approximation (in case of analytical metric) or interpolation (in case of the numerical metric) creep in. The same feature is present in both the cases, indicating the robustness of the feature.

This can be compared with the results obtained by Cunha et al in~\cite{Cunha2016}. They looked at the light ring size and the black hole shadow of both non-rotating and rotating EdGB holes. They found that for a change of $\sim4\%$ in the light ring size, the shadow size change is much smaller than expected ($\sim 1\%$). 
Here, we find that the redshift at ISCO changes by $<1\%$, despite a $\sim7\%$ change in the ISCO radius, and consequently the change in iron line is much smaller than what would be expected.

\section{Conclusion\label{sec:conclusion}}
Einstein-dilaton-Gauss-Bonnet gravity is one of the most interesting models of modified gravity. Black holes of both non-rotating and rotating type exist and are known, albeit in numerical form, making it one of the most promising alternative theories that can be tested with observational data. An analytical approximation for the numerical black holes of EdGB gravity was presented in~\cite{Kokkotas2017}, which allows faster computation and more detailed analysis compared the numerical metrics. In this paper we calculate the innermost stable circular orbits and the shape of relativistically broadened iron lines to test the accuracy of the analytical metric in the context of X-ray reflection spectroscopy. We find that the analytical metric is able to calculate the ISCO radius within $1\%$ accuracy and the iron line is calculated with an accuracy of a few percent up to $p=0.97$, which is the limit of the approximation.

In addition to being accurate, the analytical metric was used to find an interesting feature of EdGB black holes. Despite the ISCO radius changing by as much as $7\%$ from $p=0$ to $p=0.97$, the redshift changes by less than $1\%$. Consequently, the iron line is much less sensitive to the dilaton charge than expected. Since it is known~\cite{Zhang2017} that rotating EdGB black holes have some unique features that distinguish them from Kerr black holes in the context of X-ray reflection spectroscopy, an analytical metric for rotating black holes and a similar analysis of the rotating black hole solutions promises to enlighten us more on this very interesting alternative theory of gravity. Work on both these fronts is currently underway.

Recently it has been shown in \cite{Konoplya:2018arm} that the rotating analytical EdGB metric can be further simplified in such a way, that a number of potentially observable properties (such as particle's binding energy or ISCO frequencies) remain almost the same. This simplified approximate metric belongs to the class for which complete separation of variable in the Klein-Gordon and Hamilton-Jacobi equations is possible. An interesting question for further study is whether the shape of relativistically broadened iron lines for such a rotating simplified metric would be close to the full numerical metric. 

\section{Acknowledgements}
S.N. acknowledges support from the Excellence Initiative at Eberhard-Karls Universit\"at, T\"ubingen. This publication has been prepared with the support of the ``RUDN University Program 5-100''.

\appendix
\section{The Innermost Stable Circular Orbit\label{app:isco}}
The ISCO radius of a static and spherically symmetric black holes is obtained by checking the radial stability of a circular orbit,
\begin{align}
	\frac{\partial^2 V_{\textrm{eff}}}{dr^2} = 0,
\label{eq:iscodef}
\end{align}
where
\begin{align}
	V_{\rm eff} = -\frac{E^2 g_{\phi\phi} + L^2_z g_{tt}}{ g_{tt} g_{\phi\phi}} - 1,
\end{align}
and $E$ and $L_z$ are the two constants of motion, the specific energy and specific angular momentum at infinity respectively, of a particle in a circular orbit on the equatorial plane. In terms of $\Omega$, defined in Eq.~\eqref{eq:omega}, and metric coefficients, these are given as
\begin{align}
	E = - \frac{g_{tt}}{\sqrt{- g_{tt} - \Omega^2 g_{\phi\phi}}}, \qquad\qquad L_z = \frac{\Omega g_{\phi\phi}}{\sqrt{- g_{tt} - \Omega^2 g_{\phi\phi}}}.
\end{align}
Eq.~\eqref{eq:iscodef} can then be rewritten as
\begin{align}
	\frac{-1}{g_{tt} +  \Omega^2 g_{\phi\phi}}\left [g_{tt}'' -2\frac{g_{tt}'}{g_{tt}} + \Omega^2\left (g_{\phi\phi}'' -2\frac{g_{\phi\phi}'}{g_{\phi\phi}} \right )\right ] = 0,
\label{eq:iscodef2}
\end{align}
where the prime denotes differentiation with respect to $r$. We multiply both sides of the above equation by $g_{tt} +  \Omega^2 g_{\phi\phi}$. Moreover, from the metric definition in Eq.~\eqref{eq:metric}, we can see that $g_{\phi\phi} = r^2$ in the equatorial plane. Thus, the above equation becomes
\begin{align}
	g_{tt}'' -2\frac{g_{tt}'}{g_{tt}} -  6\,\Omega^2= 0.
\label{eq:iscodef3}
\end{align}
Further, using Eq.~\eqref{eq:omega} with $g_{\phi\phi} = r^2$, $\Omega$ becomes
\begin{align}
	\Omega = \pm \sqrt{- \left( \frac{\partial_r g_{tt}}{2r}\right)},
\label{eq:omega2}
\end{align}
and after substituting this in Eq.~\eqref{eq:iscodef3}, we get
\begin{align}
	g_{tt}''  - 2\frac{g_{tt}'}{g_{tt}} + 3\frac{g_{tt}'}{r} = 0.
\label{eq:iscodef4}
\end{align}
The metric component $g_{tt}$, obtained via the third order expansion described earlier, becomes
\begin{align}
	g_{tt} \approx -\left(1-\frac{r_0}{r}\right)\frac{{\cal N}_1}{{\cal D}_1},
\end{align}
where
\begin{eqnarray}
	{\cal N}_1&=& 30888 r r_0(r+r_0)(927r-1060r_0)p^6-3r_0(145693952r^3- 24067680r^2r_0 -156948260 r r_0^2-5338905r_0^3)p^5\nonumber\\
 	&+& (3750946056r^4-3062334104r^3r_0-325162656r^2r_0^2 - 1478746401 r r_0^3 -53126788r_0^4)p^4 \nonumber \\
	 &-& 2(6293682780r^4-7334803204r^3r_0-306613944r^2r_0^2 - 934415049 r r_0^3 +61245382r_0^4)p^3 \nonumber \\
 	&+& 8(1350407212r^4-2160940683r^3r_0-64904931r^2r_0^2 - 139116640 r r_0^3 +62251200r_0^4))p^2\nonumber \\
 	&+& 1048(1846581r^4 + 3798205r^3r_0 + 155610r^2r_0^2 + 270655 r r_0^3-321860r_0^4)p - 7666120r^3(509r-275r_0), \\
{\cal D}_1&=&11528 (1-p) (5-3 p) r^3 \left[ 117(927r-1060r_0)p^2-(74741r-121424r_0)p-67697r+36575r_0\right].
\end{eqnarray} 
This expression for the metric coefficient can be substituted in Eq.~\eqref{eq:iscodef4} and solved for $r$ to get the ISCO radius. 

To see how the ISCO radius changes with $p$ analytically, we can expand Eq.~\eqref{eq:iscodef4} around the Schwarzschild values, i.e., $p=0$. The computation is straightforward but tedious and we use Mathematica\textregistered\;to expand this equation up to first order in $r$ and $p$. While doing this, we have two options, either set $r_0=1$ (in which case $r_\textrm{ISCO}$ is in units of $r_0$), or set the ADM (Arnowitt-Deser-Misner) mass $M=1$ (in which case $r_\textrm{ISCO}$ is in units of $M$). In the former case, $r_\textrm{ISCO}$ in the units of $r_0$ is given as
\begin{align}
	r_\textrm{ISCO} = 3 + \frac{695651545}{4047687204}\,p + \mathcal{O}(p^2),
\label{eq:riscorh_delp}
\end{align}
while in the latter case, $r_0$ can be written, using Eq.~\ref{eq:epsilon} and the fact that $\epsilon$ defines the relation between the event horizon and the asymptotic mass $M$,
\begin{align}
	\epsilon = - \left (1-\frac{2M}{r_0} \right ),
\end{align}
as
\begin{align}
	r_0 = \frac{2M}{1+p/11-p^2/131}.
\label{eq:rhorizon}
\end{align}
Thus, $r_\textrm{ISCO}$ in the units of $M$ is given as
\begin{align}
	r_\textrm{ISCO} = 6 - \frac{16773601476568}{90201435572259}\,p + \mathcal{O}(p^2).
\label{eq:riscom_delp}
\end{align}
\bibliography{references}
\end{document}